\documentclass[12pt,article]{iopart}
\usepackage{epsfig}

\begin{document}


\title{ Nambu Monopoles in Lattice Electroweak Theory}
\vspace{1cm}

\author{B.L.G.Bakker}
 \address{ Department of Physics and Astronomy, Vrije Universiteit,
Amsterdam, The Netherlands.} \ead{blg.bakker@few.vu.nl}
\author{A.I.Veselov}
 \address{ ITEP, B.Cheremushkinskaya 25, Moscow, 117259, Russia.}
 \ead{veselov@itep.ru}
\author{M.A.Zubkov}
 \address{ ITEP, B.Cheremushkinskaya 25, Moscow, 117259, Russia.}
\ead{zubkov@itep.ru}
\date{\today}

\begin{abstract}
We considered the lattice electroweak theory at realistic values of
$\alpha$ and $\theta_W$ and for large values of the Higgs mass. We
investigated numerically the properties of topological objects that are
identified with quantum Nambu monopoles. We have found that the action
density near the Nambu monopole worldlines exceeds the density averaged
over the lattice in the physical region of the phase diagram. Moreover,
their percolation probability is found to be an order parameter for the
transition between the symmetric and the broken phases. Therefore, these
monopoles indeed appear as real physical objects. However, we have found
that their density on the lattice increases with increasing ultraviolet
cutoff. Thus we conclude, that the conventional lattice electroweak
theory is not able to predict the density of Nambu monopoles. This means
that the description of Nambu monopole physics based on the lattice
Weinberg - Salam model with finite ultraviolet cutoff is incomplete. We
expect that the correct description may be obtained only within the
lattice theory that involves the description of TeV - scale physics.
\end{abstract}

\maketitle

\section{Introduction}
The electroweak theory does not contain topologically stable monopole-like
 objects. However, certain unstable objects of topological nature
 still exist in this theory. One of the examples is the so-called
Nambu monopole\cite{Nambu}.  It must be connected by the so-called $Z$
 string with the corresponding antimonopole. The $Z$ string has nonzero
 tension. Therefore only the monopole - antimonopole  bound state may
appear as an observable object. The mass of the Nambu monopole (realized
as a classical field configuration) was estimated to be of the order of
 several TeV.  This is not far from the energies that may be achieved
by modern colliders, in particular, the LHC. Thus one may suppose,
that an indication of its existence may be detected in the near future.

Nambu monopoles are not described by means of a perturbation expansion
around the trivial vacuum background. Therefore, nonperturbative
methods should be used in order to investigate their physics. Lattice
methods seem to be one of the ways to  deal with Nambu monopoles. It
should be stressed that the mass of the Nambu monopole is close to the
energy scale, where (as commonly believed) the Standard Model does not
work\cite{TEV,Extention}. This creates an additional difficulty while
considering the problem.

Qualitative lattice investigations of the properties of Nambu monopoles
in the Standard Model have been performed both at zero and finite
temperature in the unphysical region of large coupling constants
\cite{BVZ2003,BVZ2004, BVZ2005, BVZ2006}. Nambu monopoles were found
to be condensed in the symmetric phase of lattice theory (and above
the electroweak transition in the finite temperature theory). In the
present paper we continue this investigation for realistic values of
the renormalized coupling constants ($\alpha \sim 1/128$ and $\theta_W =
\pi/6$) within the zero temperature theory. It should be stressed that
originally Nambu monopoles were defined as classical objects\cite{Nambu}.
Therefore there could appear several lattice definitions of {\it quantum}
Nambu monopoles. In this paper we discuss two of them and investigate
the difference between the positions of the corresponding monopole
trajectories.

\begin{figure}
\begin{center}
 \epsfig{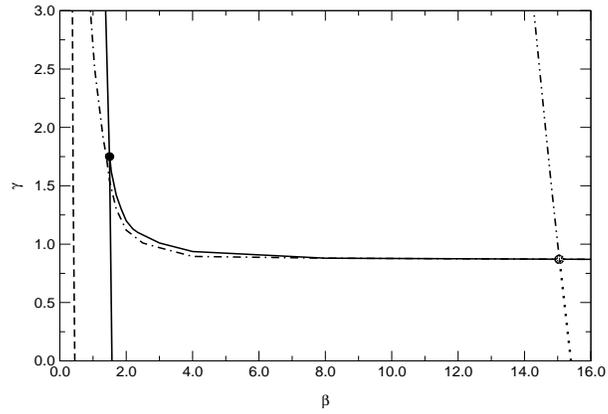}
\caption{\label{fig.1} The phase diagrams of the models in the
 $(\beta, \gamma)$-plane. The dashed line and the dot-dashed line
represent phase transitions in the $SU(2)\otimes U(1)$ model. The transitions
for the $SU(2)\otimes U(1)/Z_2$-symmetric model are represented by the solid
lines. The double-dotted-dashed vertical line on the right-hand side of the
diagram represents the line, where the renormalized $\alpha$ is constant and
equal to $1/128$.  }
\end{center}
\end{figure}

The numerical investigation of Gauge-Higgs models has a long history.
 First studies of nonabelian Gauge-Higgs systems were performed in
 eighties and were devoted mainly to  the investigation of $SU(2)$
 Gauge-Higgs model at zero temperature  (see, for example,
 \cite{Jersak,Montvay,12,13,14} and references therein). In these
studies the general phase structure of the model was established. At small
values of the scalar self coupling $\lambda$ the first order phase transition
between the physical Higgs phase and the unphysical symmetric phase of the
model was found. The phase transition becomes weaker as $\lambda$ is increased.
Using the weak coupling expansion \cite{phi4} it was found that the maximal
value of the ultraviolet cutoff in the theory is achieved at infinite
$\lambda$. At $\lambda \rightarrow \infty$ also the absolute upper bound on the
Higgs boson mass is achieved which was found to be of the order of $10 M_W$
(see, for example, \cite{12,13,14}). An early study of the $SU(2)\otimes U(1)$
Gauge-Higgs model was performed in \cite{SU2U1}, where the phase diagram of the
model was given. In general, the description of the phase diagram is similar to
that of given in the present paper (see Fig. \ref{fig.1}, and discussion in
section $5.1$).

The next step in the numerical investigation of Gauge-Higgs models was
motivated by an attempt to explore the physics of finite temperature
electroweak phase transition and its relation to cosmology, in
particular, to the problem of baryon asymmetry (see, for example,
\cite{1,2,3,4,5,6,7,8,9,10,11,EW_T} and references therein). One of
the achievements of these studies was that the endpoint of the phase
transition line (in the $T - M_H$ plane) was found. It was found that
at the experimentally allowed values of the Higgs mass the transition
in the standard electroweak theory is a crossover.

In $4D$ lattice studies of $SU(2)$ Gauge-Higgs theory the scale is fixed by the
value of the $W$-boson mass. Namely, the mass measured in lattice units is $M_W
= a \, \times \, 80 \,{\rm Gev}$, where $a$ is the value of the lattice
spacing. The ultraviolet cutoff is $\Lambda = \frac{1}{a}$. In Table
\ref{Table} we summarize the data on the ultraviolet cutoff used in selected
lattice studies of the $SU(2)$ Gauge-Higgs model. In all these studies the
coupling constant corresponding to the fine structure constant of the
Weinberg-Salam model is around $\alpha \sim 1/100$. The potential for the Higgs
boson was used with different values of $\lambda$.

\begin{table}
\label{Table} \caption{The values of the cutoff used in some selected lattice
studies of the $SU(2)$ Gauge - Higgs model.}
\begin{center}
\begin{tabular}{|c|c|c|}
\hline
{\bf Reference}  & {\bf ultraviolet Cutoff} (GeV) & {\bf $M_H$} (GeV)\\
\hline
\cite{1}  & 140 (space direction) 570 (time direction) & 80 \\
\hline \cite{2}  & 280 (time direction) & 80 \\
\hline \cite{3}  & 280 & 34 \\
\hline \cite{4}  & 110 & 16 \\
\hline \cite{5}  & 90 (space direction) 350 (time direction) & 34 \\
\hline \cite{6}  & 280 & 48 \\
\hline \cite{7}  & 140 & 35 \\
\hline \cite{8}  & 280 & 20 , 50 \\
\hline \cite{9}  & 190 & 50 \\
\hline \cite{10}  & 260 & 57 - 85 \\
\hline \cite{11}  & 200 - 300 & 47 - 108 \\
\hline \cite{12}  & 400 & 480 \\
\hline \cite{13}  & 330 -  470 & 280 - 720
(both $\lambda =\infty$ and finite $\lambda$) \\
\hline \cite{14}  & 250 -  470 &  720
(both $\lambda =\infty$ and finite $\lambda$) \\
\hline
\end{tabular}
\end{center}
\end{table}

The numerical investigation of the monopole-like \cite{U1monopole} and string -
like topological objects
  in Abelian Gauge-Higgs systems has also a long
history (see, for example, \cite{Ranft}). For an early study of the percolation
of monopole currents see \cite{Mperc}. It has been found that the monopole
density drops sharply to zero in the Coulomb phase of the model while in the
confining phase it is nonzero. The vortex density is nonzero in the Coulomb
phase and drops to zero in the Higgs phase.  The situation with topological
defects in nonabelian Gauge-Higgs systems is in general similar to that of the
Abelian ones. The first numerical
 investigation of topological defects in $3D$ lattice $SU(2)$
Gauge-Higgs model (corresponding to finite temperatures) was performed
in \cite{Chernodub}. It was found that the percolation of $Z$-vortices
is the order parameter for the electroweak transition. The conjecture
that Nambu-monopole percolation feels the electroweak transition was
made first in \cite{Chernodub_Nambu}.

In our earlier papers we considered the appearance of an
additional discrete symmetry in the fermion sector of the Standard
Model\cite{BVZ2003,BVZ2004, BVZ2005, BVZ2006}. This additional symmetry
 allows to define the Standard Model with the gauge group $SU(3)\times
SU(2) \times U(1)/{\cal Z}$, where ${\cal Z}$ is equal to $Z_6$, or
to one of its subgroups: $Z_3$ or $Z_2$. It is worth mentioning that
it has been recognized much earlier that the Standard Model appears
with the gauge group $SU(3)\times SU(2) \times U(1)/Z_6$ as a result
of the spontaneous breakdown in the $SU(5)$ unified model\cite{Z6}.
Independently, the $Z_6$ symmetry in the Higgs sector of the Standard
Model was considered in \cite{Z6f}.

In the present paper we use two lattice realizations of the electroweak
theory: with the gauge groups $SU(2) \times U(1)/Z_2$, and $SU(2)
\times U(1)$, respectively. The $SU(2) \times U(1)/Z_2$ model should be
the part of the Standard Model with the gauge group $SU(3)\times SU(2)
\times U(1)/Z_2$ or $SU(3)\times SU(2) \times U(1)/Z_6$, while the
$SU(2) \times U(1)$ model could be the part of the Standard Model with
the gauge group $SU(3)\times SU(2) \times U(1)/Z_3$ or $SU(3)\times
SU(2) \times U(1)$. We comment on the difference between the two
lattice models, which appears in our numerical research in the
unphysical region with a large coupling constant.

The paper is organized as follows.  In Sect.~$2$ we describe the lattice
models under consideration. In Sect.~$3$ we present the definition and
the main properties of quantum Nambu monopoles. Sect.~$4$ contains our
description of the quantities to be measured. In Sect.~$5$ we report our
main numerical results, while in Sect.~$6$ we discuss the difference
between the two versions of the lattice electroweak model. The final
section contains our conclusions.

\section{Lattice models under investigation}

In this section we describe the lattice models under consideration. We
do not consider the color sector of the Standard Model. Therefore, we
are left with two possibilities: the gauge groups $SU(2) \times
U(1)/Z_2$, and $SU(2) \times U(1)$. We also neglect dynamical fermions.
In both cases we use the following lattice variables:

1. The gauge field ${\cal U} = (U, \theta)$, where
\begin{eqnarray}
 \quad U = \left( \begin{array}{c c}
 U^{11} & U^{12}  \\
 -[U^{12}]^* & [U^{11}]^*
 \end{array}\right)
 \in SU(2), \quad e^{i\theta} \in U(1),
\end{eqnarray}
are realized as link variables.

2. A scalar doublet
\begin{equation}
 \Phi_{\alpha}, \;\alpha = 1,2.
\end{equation}

The potential for the scalar field is considered in its simplest form
\cite{BVZ2004} in the London limit, i.e., in the limit of infinite bare Higgs
mass. In the lattice study this does not mean, however, that the physical Higgs
mass is infinite\cite{Montvay}. Instead we expect only that it should not be
less than the inverse lattice spacing. This is indeed confirmed via direct
calculation. From the very beginning we fix the unitary gauge $\Phi_1 =
\sqrt{\gamma}$, $\Phi_2 = 0$.

For the case of the $SU(2) \times U(1)/Z_2$ symmetric model we chose the
action of the form
\begin{eqnarray}
 S_g & = & \beta \!\! \sum_{\rm plaquettes}\!\!
 ((1-\mbox{${\small \frac{1}{2}}$} \, {\rm Tr}\, U_p \cos \theta_p)
 + \mbox{${\small \frac{1}{2}}$} (1-\cos 2\theta_p))+\nonumber\\
 && + \gamma \sum_{xy}(1 - Re(U^{11}_{xy} e^{i\theta_{xy}})),
\end{eqnarray}
where the plaquette variables are defined as $U_p = U_{xy} U_{yz}
U_{wz}^* U_{xw}^*$, and $\theta_p = \theta_{xy} + \theta_{yz} -
\theta_{wz} - \theta_{xw}$ for the plaquette composed of the vertices
$x,y,z,w$.

For the case of the conventional $SU(2) \times U(1)$ symmetric model we
use the action
\begin{eqnarray}
 S_g & = & \beta \!\! \sum_{\rm plaquettes}\!\!
 ((1-\mbox{${\small \frac{1}{2}}$} \, {\rm Tr}\, U_p )
 + 3 (1-\cos \theta_p))+\nonumber\\
 && + \gamma \sum_{xy}(1 - Re(U^{11}_{xy} e^{i\theta_{xy}})).
\end{eqnarray}
In both cases the bare Weinberg angle is $\theta_W = \pi/6$, which is close to
its experimental value. The renormalized Weinberg angle is to be calculated
through the ratio of the lattice masses:  ${\rm cos} \, \theta_W = M_W/M_Z$.
The bare fine structure constant $\alpha$ is expressed through $\beta$ as
$\alpha = 1/4 \pi \beta$.   However, the renormalized coupling extracted from
the potential for infinitely heavy fermions differs from this simple
expression, as will be shown in the next sections. The physical meaning of the
constant $\gamma$ is that it is equal to the square of the vacuum value of
$\Phi_1$.

The following variables are considered as creating a $Z$ boson and a
$W$ boson, respectively:
\begin{eqnarray}
  Z_{xy} & = & Z^{\mu}_{x} \;
 = {\rm sin} \,[{\rm Arg} U_{xy}^{11} + \theta_{xy}],
\nonumber\\
 W_{xy} & = & W^{\mu}_{x} \,= \,U_{xy}^{12} e^{-i\theta_{xy}}.
\end{eqnarray}
Here, $\mu$ represents the direction $(xy)$.

After fixing the unitary gauge the electromagnetic $U(1)$ symmetry
remains:
\begin{eqnarray}
 U_{xy} & \rightarrow & g^\dag_x U_{xy} g_y, \nonumber\\
 \theta_{xy} & \rightarrow & \theta_{xy} -  \alpha_y/2 + \alpha_x/2,
\end{eqnarray}
where $g_x = {\rm diag} (e^{i\alpha_x/2},e^{-i\alpha_x/2})$.

In the unitary gauge there is also a $U(1)$ lattice gauge field, which
is defined as
\begin{equation}
 A_{xy}  =  A^{\mu}_{x} \;
 = \,[-{\rm Arg} U_{xy}^{11} + \theta_{xy}]  \,{\rm mod} \,2\pi,
\label{A}
\end{equation}
The fields $A$, $Z$, and $W$ transform as follows:
\begin{eqnarray}
 A_{xy} & \rightarrow & A_{xy} - \alpha_y + \alpha_x, \nonumber\\
 Z_{xy} & \rightarrow & Z_{xy}, \nonumber\\
 W_{xy} & \rightarrow & W_{xy}e^{-i\alpha_x}.
\label{T}
\end{eqnarray}
It should be mentioned that the field $A$ cannot be treated as the
usual electromagnetic field, because the set of variables $A$, $Z$, and
$W$ do not diagonalize the kinetic part of the pure gauge action in its
naive continuum limit.  In our lattice model the electromagnetic field
$A_{\rm EM}$ should be defined as
\begin{equation}
 A_{\rm EM}  =  A + Z^{\prime} - 2 \,{\rm sin}^2\, \theta_W Z^{\prime},
\label{A_em}
\end{equation}
where $Z^{\prime} = [ {\rm Arg} U_{xy}^{11} + \theta_{xy}]{\rm mod}
2\pi$.

\section{Nambu monopoles}

First, we define the continuum electroweak fields as they appear in the
Weinberg-Salam model in the way appropriate for the topological
consideration. Namely, after fixing the unitary gauge $\Phi_1=const.$,
$\Phi_2 = 0$, where $\Phi$ is the scalar field of the electroweak
theory, the $Z$-boson field $Z^{\mu}$ and electromagnetic field $A_{\rm
EM}^{\mu}$ are defined as
\begin{eqnarray}
 Z^{\mu} = \frac{1}{2}{\rm Tr}C^{\mu}\sigma_3 +  B^{\mu},
\nonumber\\
 A_{\rm EM}^{\mu} =  2 B^{\mu}  - 2 \,{\rm sin}^2\, \theta_W Z^{\mu},
\end{eqnarray}
where $C^{\mu}$ and $B^{\mu}$ are the corresponding $SU(2)$ and $U(1)$
gauge fields of the Standard Model.

Nambu monopoles are defined as the endpoints of the so-called
$Z$-string \cite{Nambu}. The $Z$-string is the classical field
configuration that represents an unstable object, which is
characterized by the magnetic flux extracted from the $Z$-boson field.
Namely, for a small contour $\cal C$ winding around the $Z$ - string
one should have
\begin{equation}
 \int_{\cal C} Z^{\mu} dx^{\mu} \sim 2\pi;\,
 \int_{\cal C} A_{\rm EM}^{\mu} dx^{\mu} \sim 0;\,
 \int_{\cal C} B^{\mu} dx^{\mu} \sim 2\pi {\rm sin}^2\, \theta_W .
\end{equation}
The string terminates at the position of the Nambu monopole. The hypercharge
flux is supposed to be conserved at that point\footnote{On the classical level
the monopole - like topological objects with nontrivial hypercharge flux have
an infinite self energy. At the quantum level such objects are present in the
lattice theory at the finite values of lattice spacing. However, in the
physically interesting region of lattice coupling constant $\beta \sim 15$ the
density of these objects vanishes (see Section $6$ of the present paper).}.
Therefore, a Nambu monopole carries electromagnetic flux $4\pi {\rm sin}^2\,
\theta_W$. The size of Nambu monopoles was estimated \cite{Nambu} to be of the
order of the inverse $Z$-boson mass, while its mass should be of the order of a
few TeV. According to \cite{Nambu} Nambu monopoles may appear only in the form
of a bound state of a monopole-antimonopole pair.

In lattice theory the classical solution corresponding to a $Z$-string
should be formed around the $2$-dimensional topological defect which
is represented by the integer-valued field defined on the dual lattice
\begin{equation}
 \Sigma = \frac{1}{2\pi}^*([d Z^{\prime}]_{{\rm mod} 2\pi} - d Z^{\prime})
\end{equation}
(Here we used the notations of differential forms on the lattice. For a
definition of those notations see, for example, ~\cite{forms}.)
Therefore, $\Sigma$ can be treated as the worldsheet of a {\it quantum}
$Z$-string\cite{Chernodub_Nambu,BVZ2006,Chernodub}.

Then, worldlines of quantum Nambu monopoles appear as the boundary of
the $Z$-string worldsheet:
\begin{equation}
 j_Z = \delta \Sigma
\label{jN}
\end{equation}
It has been mentioned in the previous section
that our lattice models
become $U(1)$ gauge models after fixing the unitary gauge. The
corresponding compact $U(1)$ gauge field is given by Eq.~(\ref{A}).
Therefore one may try to extract monopole trajectories directly from
$A$. Actually this was done in our earlier papers
\cite{BVZ2003,BVZ2004, BVZ2005, BVZ2006}. The monopole current is given
by
\begin{equation}
 j_{A} = \frac{1}{2\pi} {}^*d([d A]{\rm mod}2\pi)
\label{Am}
\end{equation}
Both $j_Z$ and $j_A$ represent objects carrying magnetic charge.
Therefore it would be instructive to reveal the correspondence between
them. We have
\begin{equation}
 A  =  [- Z^{\prime} + 2 \theta]{\rm mod}2\pi .
\end{equation}
In continuum notation this would be
\begin{equation}
 A^{\mu}  =  - Z^{\mu} + 2 B^{\mu},
\end{equation}
where $B$ is the hypercharge field. Its strength is divergenceless. As
a result in continuum theory the net $Z$ flux emanating from the center
 of the monopole is equal to the net $A$ flux with the opposite sign.
(Both $A$ and $Z$ are undefined inside the monopole.)  This means that
in the continuum limit the position of the Nambu monopole must coincide
 with the position of the antimonopole extracted from the field $A$.
Therefore, one can consider Eq.~(\ref{Am}) as another definition of a
quantum Nambu monopole. It is interesting that the definition (\ref{Am})
is not directly related to any observable string, as the Dirac string
connecting the corresponding lattice monopoles is invisible.

\section{Quantities to be measured}

\subsection{Evaluation of the lattice spacing}

The physical scale is given in our lattice theory by the value of the
$Z$-boson mass $M^{\rm phys}_Z \sim 90$ GeV. Therefore the lattice
spacing is evaluated to be $a \sim [90\,{\rm GeV}]^{-1} M_Z$, where $M_Z$
is the $Z$ boson mass in lattice units.

In order to evaluate the mass of the $Z$-boson we use the correlator
\cite{Montvay}:
\begin{equation}
 \sum_{\bar{x},\bar{y}}\langle \sum_{\mu} Z^{\mu}_{x} Z^{\mu}_{y} \rangle  \sim
  e^{-M_{Z}|x_0-y_0|} + e^{-M_{Z}(L - |x_0-y_0|)}  ,
\label{cor}
\end{equation}
Here the summation $\sum_{\bar{x},\bar{y}}$ is over the three ``space"
components of the four - vectors $x$ and $y$ while $x_0, y_0$ denote their
 ``time" components. $L$ is the lattice length in the ``time" direction.

It is worth mentioning, that in the $Z$-boson channel many photon states also
exist. The mass of the corresponding state on the finite lattice we used is,
however, larger than that of the $Z$ - boson itself.  For example, on the
lattice $16^3\times 24$ the minimal mass of the $3$ - photon state is $M_{3
\gamma} = 2\frac{2\pi}{16}+\frac{4\pi}{16} \sim 1.5$. Moreover, from the point
of view of perturbation theory this state appears in the correlator (\ref{cor})
through a virtual loop and is suppressed by the factor $\alpha^3$.

\subsection{The Higgs boson mass.}

The Higgs boson mass in lattice units is measured using the correlator
\begin{equation}
  \sum_{\bar{x},\bar{y}}(\langle H_{x} H_{y}\rangle - \langle H \rangle^2)
  \sim
  e^{-M_{H}|x_0-y_0|}+ e^{-M_{H}(L - |x_0-y_0|)},
\label{corH}
\end{equation}
where $H$ is the Higgs boson creation operator.

We used three different operators that create Higgs bosons:
\begin{equation}
H_x = \sum_{y} |W_{xy}|^2,\label{HW}
\end{equation}

\begin{equation}
H_x = \sum_{y} Z^2_{xy}\label{HZ}
\end{equation}
and
\begin{equation}
H_x = \sum_{y} Re(U^{11}_{xy} e^{i\theta_{xy}})
\end{equation}

Here $H_x$ is defined at the site $x$, the sum $\sum_y$ is over its neighboring
sites $y$.

\subsection{The renormalized coupling}

The bare constant $\alpha = e^2/4\pi$ (where $e$ is the electric
charge) can be easily calculated in our lattice model. It is found to
be equal to $1/(4\pi \beta)$. Therefore, its physical value
$\alpha(M_Z)\sim 1/128$ could be achieved at values of $\beta$ in the
vicinity of $10$. This naive guess is, however, to be corrected by the
calculation of the renormalized coupling constant $\alpha_R$. We
perform this calculation using the potential for infinitely heavy
external fermions. We consider Wilson loops for the right-handed
external leptons:
\begin{equation}
 {\cal W}^{\rm R}_{\rm lept}(l)  =
 \langle {\rm Re} \,\Pi_{(xy) \in l} e^{2i\theta_{xy}}\rangle.
\label{WR}
\end{equation}
Here $l$ denotes a closed contour on the lattice. We consider the
following quantity constructed from the rectangular Wilson loop of size
$r\times t$:
\begin{equation}
 {\cal V}(r) = {\rm log}\,\lim_{t \rightarrow \infty}
 \frac{  {\cal W}(r\times t)}{{\cal W}(r\times (t+1))}.
\end{equation}
Owing to the exchange of virtual photons at large distances we expect the
appearance of the Coulomb interaction
\begin{equation}
 {\cal V}(r) = -\frac{\alpha_R}{r} + const. \label{V1}
\end{equation}
It should be mentioned here, that in order to extract the renormalized value of
$\alpha$ one may apply to $\cal V$ the fit obtained using the Coulomb
interaction in momentum space. The lattice Fourier transform then gives
\begin{eqnarray}
 {\cal V}(r) & = & -\alpha_R \, {\cal U}(r)+ const,\,
\nonumber\\
{\cal U}(r) & = & \frac{ \pi}{L^3}\sum_{\bar{p}\ne 0} \frac{e^{i p_3 r}}{{\rm
sin}^2 p_1/2 + {\rm sin}^2 p_2/2 + {\rm sin}^2
 p_3/2}
 \label{V2}
\end{eqnarray}
Here $L$ is the lattice size, $p_i = \frac{2\pi}{L} k_i, k_i = 0, ..., L-1$. On
large enough lattices at $r << L$ both definitions approach each other. For
example, for $L = 75, r \in [1,10]$ the linear fit to the dependence ${\cal
U}(r)$  on $\frac{1}{r}$ gives ${\cal U}(r) \sim 0.97/r - 0.18$ while for $L =
100, r \in [1,10]$ the fit is ${\cal U}(r) \sim 0.997/r - 0.155$. However, on
lattices of the sizes we used the difference is important. Say, on the lattice
$24^4$ the fit is ${\cal U}(r) \sim 0.82/r - 0.35$ (for $r\in [1,5]$). Thus,
the values of the renormalized $\alpha_R$ extracted from  (\ref{V1}) and
(\ref{V2}) are significantly different from each other. Any of the two ways,
(\ref{V1}) or (\ref{V2}), may be considered as the {\it definition} of the
renormalized $\alpha$ on the finite lattice. And there is no particular reason
to prefer the potential defined using the lattice Fourier transform of the
Coulomb law in momentum space. Actually, our study shows that the single $1/r$
fit approximates $\cal V$ much better. Therefore, we used it to extract
$\alpha_R$. This should be compared with the results of \cite{14}, where for
similar reasons the single $e^{-\mu r}/r$ fit (instead of the lattice Yukawa
fit) was used in order to determine the renormalized coupling constant in the
$SU(2)$ Gauge-Higgs model. However, the fact that both definitions give values
that differ from each other shows the limitation on the interpretation of our
results and the importance of finite volume effects in our research.

\subsection{Nambu monopole density and percolation probability}

According to Eqs.~(\ref{jN}, \ref{Am}) the worldlines of the quantum
Nambu monopoles could be extracted from the field configurations in two
ways:
\begin{equation}
 j_Z = \delta \Sigma = \frac{1}{2\pi} {}^*d([d Z^{\prime}]{\rm mod}2\pi)
\end{equation}
and
\begin{equation}
 j_A = \delta \Sigma = \frac{1}{2\pi} {}^*d([d A]{\rm mod}2\pi).
\end{equation}
The monopole density is defined as
\begin{equation}
 \rho = \left\langle \frac{\sum_{\rm links}|j_{\rm link}|}{4L^4}
 \right\rangle,
\label{rho}
\end{equation}
where $L$ is the lattice size (in lattice units).

In order to investigate the condensation of  monopoles we use the
percolation probability $\Pi(A)$. It is the probability that two
infinitely distant points are connected by a monopole cluster (for more
details of the definition see, for example, \cite{BVZ1999}).

Both $-j_A$ and $+j_Z$ describe the same physical object. However, this
object may have a size that is larger than one lattice spacing. That's
why the two different ways to extract the monopole trajectory may give
different currents. The difference between the two currents is $j_Z -
(-j_A) = j_Z + j_A$. Therefore, the density of $j_A+j_Z$ measures the
degree of how $j_A$ differs from $-j_Z$.  In order to investigate the
difference between the two definitions of Nambu monopole currents we
use the quantity $\rho(j_A+j_Z)$, that is constructed using the current
$j_Z+j_A$ as in (\ref{rho}).

\subsection{Action density near monopole trajectories}

The monopole worldline lives on the dual lattice. Each point of the
worldline is surrounded by a three - dimensional hypercube of the
original lattice. We measure the plaquette part of the action
$S^{\rm mon}_{p}$ on the plaquettes that belong to those three-dimensional
hypercubes (normalized by the number of such plaquettes). The excess of
the plaquette action near monopole worldlines over the mean plaquette
part of the action $S_p$ is denoted by
\begin{equation}
\Delta S_p = \frac{1}{S_p}(S^{\rm mon}_{p} - S_p).
\end{equation}
Very roughly $\Delta S_p$ can be considered as measuring the magnetic
energy (both $SU(2)$ and $U(1)$), which is carried by Nambu monopoles.

We also measure $S^{\rm mon}_{l}$, which is the part of the action
$S^{\rm mon}_{l}$ on the links of the original lattice that connect
vertices of the two incident three-dimensional hypercubes mentioned
above. The excess of this link action near monopole worldlines over the
mean link part of the action $S_l$ is denoted by
\begin{equation}
\Delta S_l = \frac{1}{S_l}(S^{\rm mon}_{l} - S_l).
\end{equation}
For the simplicity of the calculations we use only one of the $8$ links
that connect incident hypercubes.

\subsection{Hypercharge monopoles}

In addition to the Nambu monopoles we also investigated the behavior of
some objects which are called hypercharge monopoles. Their worldlines
are extracted from the hypercharge field $\theta$ in the following
way:
\begin{equation}
 j_Y = \frac{1}{2\pi} {}^*d([d 2\theta]{\rm mod}2\pi)
\label{Hyp}
\end{equation}
We also define their density according to the expression (\ref{rho}).

Actually, in the naive continuum limit hypercharge monopoles would have
an infinite energy. They may appear only if one takes into account the
finiteness of the ultraviolet cutoff. It occurs that they are mainly of
interest in the strong coupling region, where the two considered
lattice models appear to behave differently.

\section{Numerical results}

\subsection{Phase diagram}

The phase diagrams of the two models under consideration are presented in Fig.
\ref{fig.1}. At small values of the coupling constants the model with
 the gauge group $SU(2)\otimes U(1)/Z_2$ has already been investigated
 in our earlier paper \cite{BVZ2004}. The model with the gauge group
 $SU(2)\otimes U(1)$ was investigated in the paper \cite{SU2U1} (also
 at small values of the coupling constants). The dashed vertical line
represents the phase transition in the $SU(2)\otimes U(1)$-symmetric model
 (we call it further Model A). This is the confinement-deconfinement
phase transition corresponding to the $U(1)$ constituents of the
model. The same transition for the $SU(2)\otimes U(1)/Z_2$-symmetric
model (we call it model B) is represented by the solid vertical line. The
dot-dashed horizontal line corresponds to the transition between the
broken and symmetric phases of model A. The solid horizontal line
represents the same transition in model B.  Interestingly, in the
$SU(2)\otimes U(1)/Z_2$ model both transition lines meet, forming a
triple point. Much attention was paid to this fact in \cite{BVZ2004}.

So, in both models there are three phases. The first one is situated
in the left-hand side of the phase diagram. In this phase there
are confinement-like forces both between the right-handed and the
left-handed external fermions. However, due to the presence of
the charged scalar field the string connecting external fermions is
broken and the confining forces disappear at a certain distance (see,
for example, \cite{BVZ2004}). In this phase both Nambu monopoles and
hypercharge monopoles are condensed. The second phase is situated below
the horizontal phase transition line and right to the vertical phase
transition line. In this phase the confining forces are observed only
between the  left-handed fermions. The hypercharge monopoles are not
condensed in this phase and their density falls sharply. For the detailed
description of different phases of the $SU(2)\otimes U(1)/Z_2$-symmetric
model (including the properties of topological objects and static
potentials) at small values of $\beta$ see \cite{BVZ2004}.

Real physics is commonly believed to be achieved within the phases of the
two models situated in the right upper corner of Fig.~$1$. In this phase
neither Nambu monopoles nor hypercharge monopoles are condensed. The
confining-like forces are not observed here both between right-handed
and left-handed fermions. The double-dotted-dashed vertical line on
the right-hand side of the diagram represents the line, where the
renormalized $\alpha$ is constant and equal to $1/128$. In order
to draw the phase-transition lines at small values of $\beta$ we
use the results of \cite{BVZ2004} and \cite{SU2U1}. These data have
been checked using the observables listed in the previous section. In
 particular, the density of hypercharge monopoles appears to be very
sensitive to the $U(1)$ confinement-deconfinement phase transition,
while the density and percolation probability of Nambu monopoles feel
the transition between the broken and symmetric phases. The position of
the horizonal line for large values of $\beta$ is obtained using mainly
the percolation probability for the Nambu monopoles (which has been found
to be an order parameter of the corresponding transition in the lattice
Standard Model, with $SU(3)$ constituents included \cite{BVZ2006}). This
position corresponds also to the maximum of the susceptibility $\chi =
\langle H^2 \rangle - \langle H\rangle^2$. (See Fig. \ref{fig.6}. We
calculated the susceptibility  using both definitions, Eqs.~(\ref{HW}
and (\ref{HZ}) for $H$.)

All simulations were performed on lattices of sizes $8^4$ and $16^4$.
Several points were checked using a lattice $24^4$. In general we found
no significant difference between the mentioned lattice sizes.

\subsection{Renormalized masses and couplings}

In the region $\beta \in (10,20)$, $\gamma \in (1,2)$ we found no
difference between the two versions of lattice electroweak theory.
Therefore, we omit mentioning to what particular model the considered
quantity belongs in this region of coupling constants.

The  $Z$-boson masses is found to change very slowly with the variation
of $\beta$.  The dependence on $\gamma$ seems to be stronger. $M_Z$
in lattice units grows with the increase of $\gamma$.

For the calculation of the Z-boson mass we used lattices of sizes:
$6^3\times 12$, $8^3\times 16$, $12^3\times 24$, and $16^3\times 24$. The
dependence of the Z-boson correlator Eq.~( \ref{cor}) on $r = x_0-y_0$
is presented in Fig.~$2$ for $\gamma = 1, \beta = 15$ and a lattice
$16^3\times 24$. From this plot we extract the mass of the $Z$-boson to
be $0.22\pm 0.02$.  It is important to notice that we did not find any
dependence of $M_Z$ on lattice size.

At $\beta = 15$ we localize the position of the transition between
the symmetric and broken phases of the model at $\gamma_c = 0.92 \pm
0.02$. The measured $Z$ - boson mass for $\gamma \in [0.9,0.94]$ is $0.21
\pm 0.02$. So, we evaluate the $Z$ - boson mass at the transition point
to be $M_Z = 0.21 \pm 0.02$. The correspondent value of the ultraviolet
cutoff (the inverse lattice spacing) is $\Lambda = 90\, {\rm GeV}/M_Z =
430 \pm 40\, {\rm GeV}$.

As for the Higgs boson mass, due to the insufficient statistics we cannot
extract $M_H$ from our data with reasonable accuracy. According to our (very
rough) estimate at  $\beta=15, \gamma = 1$ we have $M_H/M_Z \sim 9\pm 2$. This
estimate is in agreement with the investigation of the $SU(2)$ Higgs model
\cite{12,13,14} performed near the transition point for the London limit of the
Higgs potential and realistic $\beta$. Actually, as in \cite{12} we made our
estimate based on the consideration of the correlator for small ``time"
separation ($ \le 3$).

It was found in \cite{14} that at larger distances a second mass
parameter close to $2 M_W$ contributes to the correlator.  In our study
the accuracy of measurements does not allow us to extract information
from the $H$-$H$ correlator for "time" separations $\ge 4$. Therefore,
we do not see the signal of the two gauge-boson bound state. However,
we do not exclude that it would appear if more statistics is collected
and ``time" separations  $\ge 4$ are explored.

In \cite{14} in order to evaluate the Higgs boson mass in this situation
the value of mass $\sim 2 M_W$ was considered as the mass of the
bound state of the two gauge bosons, and only the first mass in the
given channel was interpreted as the Higgs boson mass. Actually, we
do not see any reason to do so, and guess that the second mass in this
channel may serve as the Higgs boson mass. However, this question must
be investigated separately. Therefore we only claim here that the Higgs
boson mass for our choice of initial parameters is larger than about
$2 M_W \sim 160$ GeV, that could be the lowest mass in the given channel.

The renormalized coupling constant $\alpha$ is found to be close to
the realistic value $\alpha(M_Z)=1/128$ along the line represented in
Fig.~$1$. In Fig.~$3$ the dependence of the potential for infinitely
heavy right-handed leptons on $1/r$ is shown for $\gamma = 1,
\beta = 15$. The renormalized $\alpha_R$ is extracted from this
dependence. Actually, a linear dependence is observed for $r\in
[1,6]$. Therefore we treat this constant as $\alpha_R(r/a) \sim
\alpha_R(M_Z)$. In Fig.~$4$ we exhibit the dependence of $1/\alpha_R$
on $\beta$ for fixed $\gamma = 1$. Here it should be mentioned that
according to the subsection IV.C of the present paper, the definition
of the renormalized $\alpha$ on the lattices of finite size suffers
from lattice artifacts.

\subsection{Nambu monopoles}

We used both definitions of Nambu monopoles given in section
$3$. In Fig.~$5$ we show their density and percolation probability as a
function of $\gamma$ along the line of constant renormalized $\alpha_R =
1/128$.  Interestingly, the density and percolation probability coincide
here for the two mentioned definitions of Nambu monopole while the precise
position of monopole trajectories differ by about $30$\%, i.e., we found
that $2\rho(j_A+j_Z)/(\rho(j_A)+\rho(j_z))\sim 0.3$. This means that the
physical Nambu monopole has a size larger than $1$ in lattice units.
Therefore the two different lattice definitions locate it sometimes
differently.

It is clear from Fig.~$5$ that the percolation probability is the order
parameter of the transition from the symmetric to the broken phase. We
did not investigated the order of the transition.  However, according
to the previous investigations of the $SU(2)$ Higgs model \cite{EW_T}
we expect that for our choice of the Higgs potential it could actually
be a crossover.

In order to compare the position of the transition between the symmetric
and broken phases with the point where the percolation probability
vanishes, we investigated the susceptibility $\chi = \langle H^2 \rangle
- \langle H\rangle^2$ extracted both from $H_Z = \sum_{y} Z^2_{xy}$
and $H_W = \sum_{y} |W_{xy}|^2$. The dependence of $\chi$ on $\gamma$
along the line of constant $\alpha = 1/128$ is shown in Fig.~$6$.

The magnetic energy $\Delta S_p$ carried by a Nambu monopole is presented
in Fig.~$7$. The excess of the link action near the monopole worldline
$\Delta S_l$ is shown in Fig.~$8$. The behavior of both variables show
that a quantum Nambu monopole may indeed be considered as a physical
object.

\subsection{Relation between the lattice model and continuum physics}

The real continuum physics should be approached along the line of constant
$\alpha_R$, i.e., along the line of constant physics (at this point we omit
consideration of $\theta_W$ and $M_H$). From our data it follows that the
ultraviolet cutoff $\Lambda = a^{-1} = (90~{\rm GeV})/M_Z$ grows with
decreasing $\gamma$ along the line of constant physics. This dependence is not
far from the tree - level estimate $\Lambda/{\rm GeV} = 90/M_Z= 90\sqrt{\beta/
\gamma}\, {\rm cos}\theta_W \sim 310/\sqrt{\gamma}$. It occurs that $\Lambda$
is increasing slowly along this line with decreasing $\gamma$ and achieves a
value close to $430 \pm 40$ GeV at the transition point between the physical
Higgs phase and the symmetric phase. (At $\beta = 15$ the transition occurs at
$\gamma_c = 0.92 \pm 0.02$, where the $Z$ boson mass in lattice units is
evaluated to be $M_Z = 0.21\pm 0.02$.) Therefore, we claim that up to finite
volume artifacts the largest achievable value of the ultraviolet cutoff is
around $430 \pm 40$ GeV if the potential for the Higgs field is considered in
the London limit.

It is interesting to understand what happens with this maximal value
of the ultraviolet cutoff if the Higgs potential would contain a finite
scalar self coupling $\lambda$ \cite{Montvay}. Then for the case of the
$SU(2) \times U(1)/Z_2$ symmetric model we chose the action of the form
\begin{eqnarray}
 S & = & \beta \!\! \sum_{\rm plaquettes}\!\!
 ((1-\mbox{${\small \frac{1}{2}}$} \, {\rm Tr}\, U_p \cos \theta_p)
 + \mbox{${\small \frac{1}{2}}$} (1-\cos 2\theta_p))+\nonumber\\
 && - \gamma \sum_{xy} Re(\Phi^+U_{xy} e^{i\theta_{xy}}\Phi) + \sum_x (|\Phi_x|^2 + \lambda(|\Phi_x|^2-1)^2),
\end{eqnarray}
where $\Phi$ is the scalar doublet. Here $\gamma = 2\kappa$, where $\kappa$
corresponds to the constant used in the investigations of the $SU(2)$ gauge
Higgs model \cite{1,2,3,4,5,6,7,8,9,10,11,12,13,14}. The weak coupling
expansion in lattice theory \cite{phi4} gives the prediction that the maximal
possible ultraviolet cutoff is achieved in lattice electroweak theory at
infinite $\lambda$. Thus  we expect that the largest achievable value of the
ultraviolet cutoff for any $\lambda$ should be close to the value  $430 \pm 40$
GeV calculated in our present work. One can compare this result with that of
the previous research (see Table $1$).

We like to note, that from the point of view of perturbation theory
the energy scale $1$ TeV appears in the so-called hierarchy problem
\cite{TEV}.  Namely, the mass parameter $\mu^2$ for the scalar field
receives a quadratically divergent contribution in one loop. Therefore,
formally the initial mass parameter ($\mu^2= - \lambda_c v^2$, where $v$
is the vacuum average of the scalar field) should be set to infinity
in such a way that the renormalized mass $\mu^2_R$ remains negative and
finite. This is the content of the so-called fine tuning. It is commonly
believed that this fine tuning is not natural \cite{TEV} and, therefore,
the finite ultraviolet cutoff $\Lambda$ should be maintained. From the
requirement that the one-loop contribution to $\mu^2$ is less than $10
|\mu_R^2|$ one derives that $\Lambda \sim 1$ TeV.

Our lattice study demonstrates the following peculiar feature of electroweak
theory. If we are moving along the line of constant $\alpha=1/128$, then the
Nambu-monopole density decreases with increasing $\gamma$ (for $\gamma
> 1$). Its behavior is approximated with a good accuracy by the simple formula:
\begin{equation}
 \rho \sim e^{2.08 - 4.6 \gamma} .\label{rho1}
\end{equation}
It is worth mentioning, that the monopole density given by (\ref{rho1})
is expressed in lattice units. The density expressed in physical units
may be defined as $\rho_c = \frac{\rho}{a^{3}}$, where $a$ is the lattice
spacing. It was mentioned above that the lattice spacing decreases with
decreasing $\gamma$. Thus the Nambu monopole density expressed in physical
units increases with decreasing $a$.

Naively one may think that the density (in lattice units) should
decrease with increasing ultraviolet cutoff and the physical value of
the density is achieved at the transition point. However, it occurs that
the situation is inverse. Thus the density in physical units does not
tend to a constant at $a \rightarrow 0$.  Therefore, it is not clear
how to extract the physical continuum density of Nambu-monopoles from the
lattice data. In this connection it is also important to notice that we
did not find any dependence of the monopole density on the lattice size.

\section{On the difference between the considered lattice models}

In the previous section we have seen that there exists no difference
between the two lattice models with the gauge groups $SU(2) \times
U(1)/Z_2$, and $SU(2) \times U(1)$ at realistic values of the coupling
constants. However, such a difference clearly exists in the region of
large coupling constants ($\alpha>0.1$), where the phase diagrams of the
two models do not coincide. In particular, this difference can be easily
seen from the behavior of the hypercharge-monopole density (see Fig.~9,
where the dependence on $\beta$  is shown for $\gamma = 1.5$). In the
same figure the density of Nambu monopoles (defined through the $Z^{\prime}$
field) is presented as well. It is clear that the two kinds of monopoles
behave differently in the two models.

The explanation of this fact may be related to the possible appearance
of the unification of fundamental interactions at the energy scale
$\Lambda$ of about $1$ TeV. Namely, it has been shown in \cite{Z2007}
that if TeV physics is described by a simply connected unified gauge
group (as in Petite Unification Models \cite{PUT,PUT1}), then the
following relation exists between the additional discrete symmetry of
the Standard Model and the monopole content of the theory, which
describes TeV physics: If the electroweak theory has the gauge group
$SU(2) \times U(1)/Z_2$, then there are topologically stable monopoles
in the unified theory, which are composed of electroweak fields (when
seen from large enough distances). From those monopoles the hypercharge
magnetic flow
\begin{equation}
 \int_{\cal C} 2B^{\mu} dx^{\mu} \sim 2\pi
\end{equation}
should emanate. Therefore, at low energies they may be identified with
the hypercharge monopoles, defined in Eq.~(\ref{Hyp}). Such objects do
not appear if the gauge group of electroweak theory is $SU(2) \times
U(1)$. Those topologically stable objects have masses of the order
of $\Lambda/\alpha$.  At realistic values of the coupling constant
they could not be observed at low energies within the electroweak
theory with the ultraviolet cutoff $\Lambda$, as their masses appear
to be much larger than the cutoff. However, if one would imagine that
the coupling constant $\alpha$ in our world becomes close to unity,
then the mass of such objects becomes comparable to $\Lambda$. If so,
their density in the case of the $SU(2) \times U(1)/Z_2$-symmetric model
must exceed considerably the same density calculated within the $SU(2)
\times U(1)$ model.

Exactly this happens in our models, where the ultraviolet cutoff $\Lambda$
is estimated to be of the order of $400$ GeV. The naive expression
$\Lambda/\alpha$ for the mass of hypercharge monopole (in the $SU(2)
\times U(1)/Z_2$-symmetric model) gives values comparable to $\Lambda$
in the region of couplings presented in the figure, where $\alpha$
is found to be of the order of $0.1$. It still remains larger than the
cutoff, but we should remember that the classical evaluation of mass
may be renormalized via quantum fluctuations. Thus, we can see that the
density of hypercharge monopoles in the $SU(2) \times U(1)/Z_2$ model
indeed exceeds the one of the $SU(2) \times U(1)$ model in the region
of small $\beta$ (the bare value of $\alpha$ is $1/(4\pi\beta)$).

\section{Conclusions}

In this paper we investigated lattice electroweak theory numerically at
realistic values of the fine structure constant and the Weinberg angle. We
considered the potential for the Higgs field in the London limit, i.e., for
infinite bare scalar self coupling.

We found that the two definitions of the theory (with the gauge groups
$SU(2)\otimes U(1)/Z_2$ and $SU(2)\otimes U(1)$, respectively) do
not lead to different predictions at these values of the couplings.
However, the corresponding models behave differently at unphysically
large values of $\alpha$. The main difference is in the behavior of the
so-called hypercharge monopoles, which would become the $Z_2$ monopoles
of the unified theory, if the latter has a simply connected gauge group.

On the phase diagram of the considered lattice model a line can be drawn,
where the renormalized fine structure constant is close to its realistic
value $\frac{1}{128}$. It should be remembered, though, that in order
to draw the true line of constant physics (where the Higgs mass and the
renormalized Weinberg angle are constant as well) one must vary the bare
Weinberg angle and the bare scalar self coupling.

Our investigation of the line of constant renormalized $\alpha$ for the
infinite bare self coupling of the Higgs field allows us to draw the conclusion
that values of lattice spacings smaller than about $(430\pm 40\, {\rm
GeV})^{-1}$ cannot be achieved in principle for this choice of the potential
for the Higgs field (at least, for the considered lattice sizes). It would be
important, therefore, to consider finite values of the scalar self couplings
and investigate finite volume artifacts in order to understand whether there is
a maximal value of the ultraviolet cutoff in the electroweak theory (that is
not related to a Landau pole in renormalized $\alpha$ and $\lambda$).

Actually, we suppose that at the point of the transition the line of constant
physics (corresponding to the Higgs phase) stops and another line of constant
physics begins (that corresponds to the unphysical symmetric phase).  Although
usually the transition is thought to be a crossover, we see that the physical
content of the theory is changed drastically at the transition.  Namely, Nambu
monopoles appear to be condensed in the symmetric phase.  So, these objects
dominate in the dynamics of this phase of the theory. This, of course,
contradicts all observable data. Thus the points of the symmetric phase cannot
be associated with real physics. The percolation probability of Nambu monopoles
appears as an order parameter for this transition. Therefore we conclude, that
the given transition may belong to the class of the transitions of the so -
called Kertesz type (see, for example, \cite{Kertesz,Kertesz2}).  That's why
one may suppose that the position of this percolation transition may not
correspond precisely to the position of the transition determined with the aid
of other physical quantities. However, according to our results, for example,
the position of the percolation transition coincides within statistical errors
with the position of the maximum of the susceptibility extracted from the Higgs
boson creation operator (see Figs. \ref{fig.5} and \ref{fig.6}).

It is worth mentioning, that the gauge boson mass does not vanish at the
transition point. So, it differs from zero in the symmetric phase of the
theory, not only in the physical Higgs phase. This is in accordance with the
previous numerical data \cite{Montvay,EW_T,1,2,3,4,5,6,7,8,9,10,11,12,13,14}.
For pure $SU(2)$ theory (the limit $\gamma \rightarrow 0$) the situation is
more complicated: The gluon propagator contains a mass parameter, but the
dependence on the momentum is not consistent with the usual mass pole (see, for
example, \cite{propagator}).

We have found that the Nambu monopole density on the lattice increases
with increasing ultraviolet cutoff. Although we obtain this result
for fixed bare Weinberg angle and infinite bare scalar self coupling,
we expect that it should remain unchanged in the full theory with
variable $\theta_W$ and $\lambda$.  Thus we conclude, that the
conventional electroweak theory is not able to predict the density of
Nambu monopoles. This is, however, not a surprise because the Standard
Model should be considered as a finite cutoff theory.  According to
common lore and in accordance with our numerical results discussed
above, the Standard Model cannot describe nature at energies above $1$
TeV. On the other hand, in \cite{Nambu} the Nambu monopole mass was (roughly)
estimated to be in the TeV region. This rough estimate, however, is obtained
based on the classical consideration of monopole configurations. In quantum
theory the density of monopoles is defined by the balance of entropy and
self energy. We found that the density of the monopoles approaches a
value of the order of $0.1$ (in lattice units) near the transition to
the unphysical phase of the theory. Therefore, it is natural to suppose
that the self energy of these objects decreases when the transition
is approached within the Weinberg - Salam model. We indeed observe the
decrease of the action near monopole trajectories (see Fig.(\ref{fig.7})
and Fig (\ref{fig.8})). So, we expect that the value of mass evaluated
near the transition should be significantly less than the classical value
calculated in \cite{Nambu}. On the other hand, it is natural to suppose
that the monopole mass approaches its classical value deep in the Higgs
phase. There the Nambu monopole currents are not observed in our study
as it should when one looks for an object with a mass above 1 TeV at
energies of the order of $100$ GeV.

To conclude, the description of Nambu monopole physics based on the lattice
Weinberg - Salam model with finite ultraviolet cutoff seems to us incomplete.
The correct nonperturbative description may, therefore, be obtained within the
lattice theory that involves the description of TeV - scale physics.
Nevertheless, on the basis of electroweak theory only we are able to reach the
conclusion that Nambu monopoles appear as real physical objects: Our numerical
results show that the action density near the Nambu monopole worldlines exceeds
the density averaged over the lattice in the physical region of the phase
diagram. This confirms the classical consideration of \cite{Nambu}, where
corresponding classical configurations were found to carry energy. Another
important piece of information about the structure of the lattice Weinberg -
Salam model comes from the consideration of the percolation of monopole
currents. The percolation probability is found to be an order parameter for the
transition between the symmetric and broken phases. In the unphysical symmetric
phase of the model Nambu monopoles are condensed, which means that this phase
of the lattice model indeed cannot be associated with the real continuum
physics.

\vspace{3ex}

A.I.V.  and M.A.Z. kindly acknowledge the hospitality of the Department of
Physics and Astronomy of the Vrije Universiteit, where part of this work
was done.  This work was partly supported by RFBR grants 05-02-16306,
07-02-00237, and 08-02-00661,  RFBR-DFG grant 06-02-04010, by Federal
Program of the Russian Ministry of Industry, Science and Technology No
40.052.1.1.1112, by Grant for leading scientific schools 843.2006.2.

\clearpage

\clearpage

\begin{figure}
\begin{center}
 \epsfig{figure=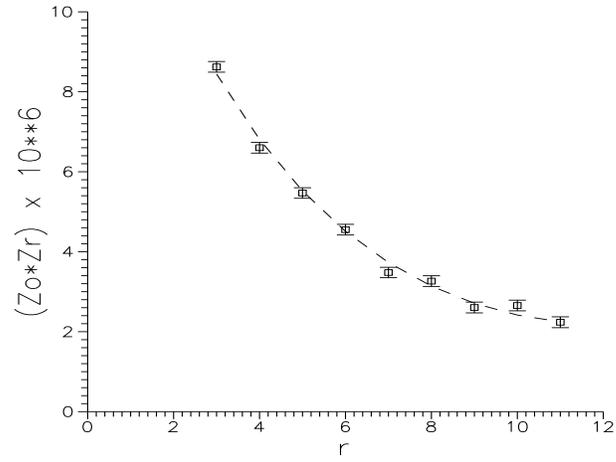,height=60mm,width=80mm,angle=0}
\caption{\label{fig.2}  $\frac{1}{4 R^6} \sum_{\bar{x}, \bar{y}} \sum_{\mu}
<Z^{\mu}_{x} Z^{\mu}_{y} >$ as a function of $r =|x_0-y_0|$. Here $R = 16$ is
the lattice size in ``space" direction.}
\end{center}
\end{figure}

\clearpage

\begin{figure}
\begin{center}
 \epsfig{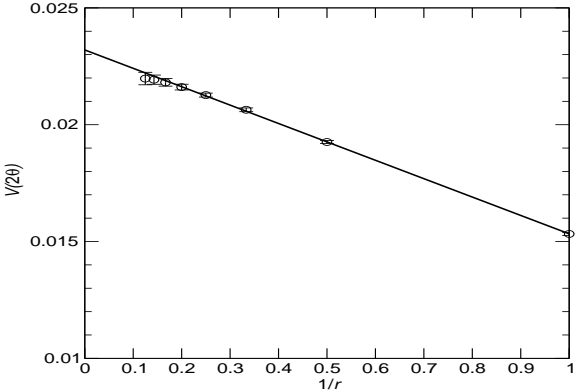}
\caption{\label{fig.3}  Potential for right-handed leptons as a function
of $1/ r$ for $\beta = 15, \gamma = 1$.}
\end{center}
\end{figure}

\clearpage

\begin{figure}
\begin{center}
 \epsfig{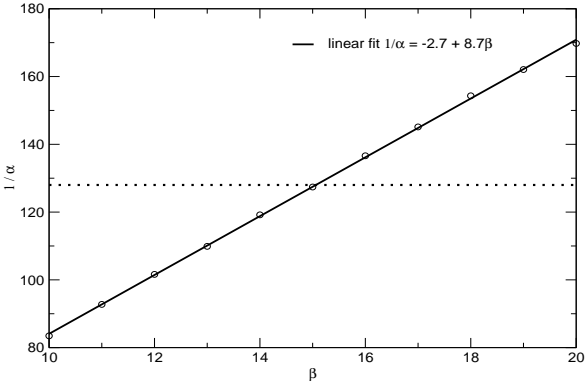}
\caption{\label{fig.4} Renormalized coupling $1/\alpha_R$ as a function
of $\beta$ for $\gamma = 1$.}
\end{center}
\end{figure}

\clearpage

\begin{figure}
\begin{center}
 \epsfig{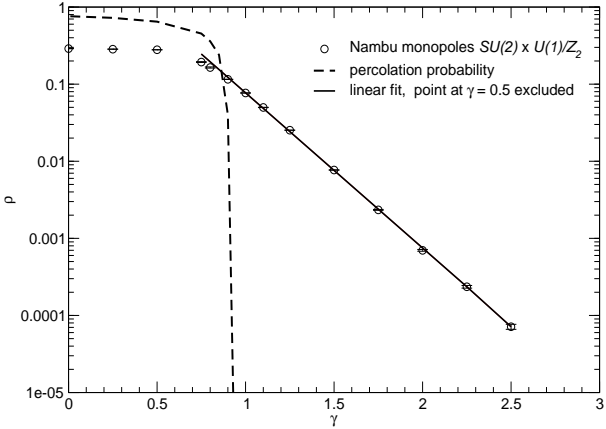}
\caption{\label{fig.5} Nambu monopole density and percolation probability as a
function of $\gamma$ along the line of constant $1/\alpha_R=128$.}
\end{center}
\end{figure}

\clearpage

\begin{figure}
\begin{center}
 \epsfig{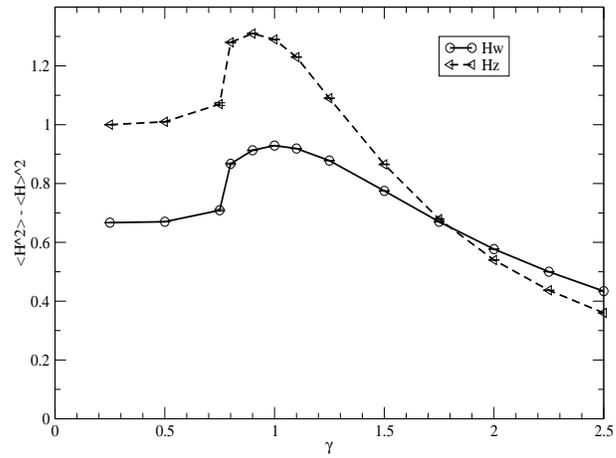}
\vspace{1.5ex}
\caption{\label{fig.6} Susceptibility $\chi = \langle H^2 \rangle -
 \langle H\rangle^2$ along the line of constant $1/\alpha_R=128$. Here $H_W$ is the operator defined in (\ref{HW}), while
 $H_Z$ is defined by (\ref{HZ}).}
\end{center}
\end{figure}

\clearpage

\begin{figure}
\begin{center}
 \epsfig{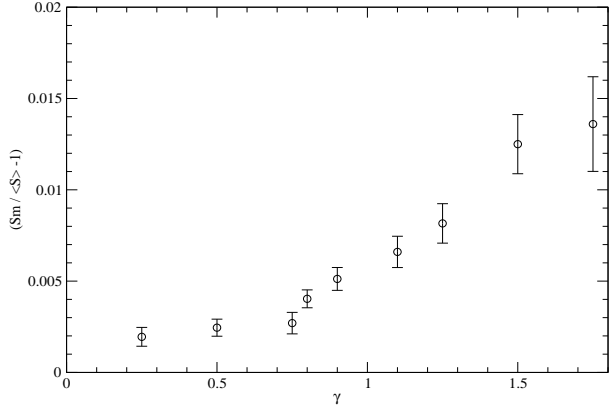}
\vspace{1.5ex} \caption{\label{fig.7} Excess of the plaquette action near
monopole trajectories along the line of constant $1/\alpha_R=128$.}
\end{center}
\end{figure}

\clearpage

\begin{figure}
\begin{center}
 \epsfig{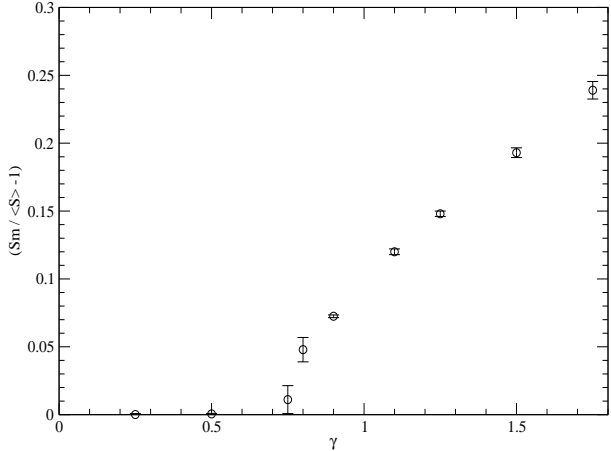}
\vspace{1.5ex}
\caption{\label{fig.8} Excess of the link action near monopole trajectories
along the line of constant $1/\alpha_R=128$.}
\end{center}
\end{figure}

\clearpage

\begin{figure}
\begin{center}
 \epsfig{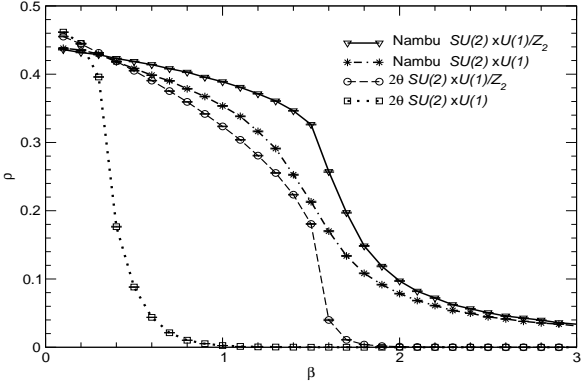}
\caption{\label{fig.10} Densities of hypercharge monopoles and Nambu
monopoles (extracted from $Z^{\prime}$) as a function of
$\beta$ for $\gamma = 1.5$ for both models.}
\end{center}
\end{figure}

\end{document}